\documentclass[twocolumn,showpacs,preprintnumbers,amsmath,amssymb,prl,aps]{revtex4}

\usepackage{epsfig}
\usepackage{graphicx}
\usepackage{dcolumn}
\usepackage{bm}

\begin{document}

\title{Understanding the bulk electronic structure of Ca$_{1-x}$Sr$_x$VO$_3$}

\author{Kalobaran Maiti,$^{1,2}$ U. Manju,$^1$ Sugata Ray,$^1$ Priya Mahadevan,$^3$ I.H.
Inoue,$^4$ C. Carbone,$^5$ and D.D. Sarma$^1$}

\altaffiliation[Also at ]{Jawaharlal Nehru Centre for Advanced
Scientific Research, Bangalore - 560 012, INDIA}

\affiliation{$^1$Solid State and Structural Chemistry Unit, Indian
Institute of Science, Bangalore 560012, INDIA\\
 $^2$Department of Condensed Matter Physics and Materials Science, Tata Institute of
Fundamental Research, Homi Bhabha Road, Colaba, Mumbai - 400 005,
INDIA\\
 $^3$Department of Physics, Indian Institute of Technology, Chennai 600036,
 INDIA\\
 $^4$Correlated Electron Research Center (CERC), AIST Tsukuba Central
4, Tsukuba 305-8562, JAPAN\\
 $^5$Istituto di Struttura della Materia, Consiglio Nazionale delle Ricerche,
Area Science Park, I-34012 Trieste, Italy}

\date{\today}

\begin{abstract}

We investigate the electronic structure of Ca$_{1-x}$Sr$_x$VO$_3$
using careful {\em state-of-the-art} experiments and calculations.
Photoemission spectra using synchrotron radiation reveal a
hitherto unnoticed polarization dependence of the photoemission
matrix elements for the surface component leading to a substantial
suppression of its intensity. Bulk spectra extracted with the help
of experimentally determined electron escape depth and estimated
suppression of surface contributions resolve outstanding puzzles
concerning the electronic structure in Ca$_{1-x}$Sr$_x$VO$_3$.

\end{abstract}


\maketitle

For several decades, Hubbard model has been the archetype to
understand a wide variety of electronic and magnetic properties in
strongly correlated electronic systems, such as transition metal
compounds. It is now well-understood that an increasing effect of
electron correlation, measured by the ratio of the intra-site
Coulomb interaction strength, $U$, and the bare bandwidth, $W$,
tends to make the system more localized in terms of an increasing
effective mass, tendencies towards eventual opening of a band gap
and the formation of a local moment. Dynamical mean field theory
(DMFT) represents one of the most successful approaches
\cite{Kotliar-RMP} to capture most of these features. However, the
parameter values of such a model Hamiltonian need to be fixed by
comparison of theoretical predictions with various experimental
results. Photoelectron spectroscopy (PES) has been extensively
used over the last two decades for this purpose. An extreme
surface sensitivity of this technique \cite{univ-curve} coupled
with the possibility of a drastically altered surface electronic
structure compared to that in the bulk may make the direct
application of PES to understand bulk properties impossible
\cite{kaindl,gabi-dd,lasrtio3,lcv,csv,lcvprb2,suga,ruth}. This has
indeed been conclusively demonstrated for certain early transition
metal oxides \cite{gabi-dd,lasrtio3,lcv,csv,lcvprb2,suga,ruth}.
Thus, it becomes necessary to devise a reliable method to separate
the bulk and the surface electronic structures before a detailed
understanding can be obtained.

In this context, Ca$_{1-x}$Sr$_x$VO$_3$ (0$\leq x \leq$1), has
firmly established itself as one of the most interesting systems,
providing a critical testing ground for the {\em state-of-the-art}
theories in the recent past
\cite{csv,suga,liebsch,andersen,isaodhva,fujimori92}. This is
primarily due to the continuous tunability of the structural
parameters arising from the fact that the V-O-V bond angle across
the series changes progressively from 180$^o$ in SrVO$_3$ to
160$^o$ in CaVO$_3$ \cite{isaoprb}. Thus, the V $d$-bandwidth,
$W$, and consequently, the correlation strength, $U/W$, are
expected to vary systematically with $x$ in
Ca$_{1-x}$Sr$_x$VO$_3$. A recent calculation based on linearized
Muffin-Tin orbital (LMTO) method within the local density
approximation (LDA) indeed established that the bandwidth changes
from about 2.8~eV for SrVO$_3$ to 2.4~eV for CaVO$_3$,
representing an impressive 14\% change in $W$ \cite{andersen}. On
the experimental side, the $\gamma$ values from specific heat
measurements \cite{isaoprb} are 6.4 and 8.6
mJ~K$^{-2}$~mole$^{-1}$ for SrVO$_3$ and CaVO$_3$ respectively,
consistent with the expected larger $U/W$ value for CaVO$_3$. In
sharp contrast, most recent photoemission results \cite{suga} have
been interpreted as suggesting almost identical electronic
structures across the series independent of the composition, $x$,
in Ca$_{1-x}$Sr$_x$VO$_3$. This unexpected result, apparently
inconsistent with more than 25\% increase in $\gamma$ and a 14\%
decrease in $W$, has naturally created a lot of interest,
prompting us to make a critical evaluation of the electronic
structure of this series of compounds. Present results establish
that there is an unexpected and unusually strong polarization
dependence of the spectral intensity of the surface component in
the experimental spectra. When this is taken into account along
with experimentally determined photoelectron escape depth,
$\lambda$, the electronic structure can be described consistently
along with the mass enhancement and the reduction in the
bandwidth, resolving the puzzling aspects of the electronic
structure of this important class of compounds.

Single crystal samples were prepared by floating zone method and
characterized by $x$-ray diffraction, Laue photography and
thermogravimetric analysis as described elsewhere \cite{isaoprb}.
The photoemission (XP) measurements were carried out on cleaved
single crystal surfaces at VUV-beamline, Elettra, Trieste, Italy
and the experimental resolution was 20-200~meV depending on the
incident photon energy of 20-800~eV. We observe that the spectra
from cleaved and scraped samples are essentially identical as
shown later in Fig. 3(d). This is not surprising since there is no
well defined cleavage plane in the perovskite structure. Thus,
cleaving often leads to a fracture along crystallographically
weaker points.

It is now well established that the surface and bulk electronic
structures are significantly different in vanadates. The
delineation of the surface and bulk contributions in the
photoemission spectra requires reliable estimates of $\lambda$,
which sensitively depends on the electron kinetic energy (KE).
Fortunately, a nearly unique aspect of the electronic structure of
this series allows us to make such estimates of $\lambda$.
V$^{4+}$ in these compounds charge disproportionates to V$^{3+}$
and V$^{5+}$ species at the surface \cite{csv}. This is
convincingly demonstrated in Fig.~1, showing the V 2$p_{3/2}$ core
level spectra at different photon energies for CaVO$_3$ and
Ca$_{0.3}$Sr$_{0.7}$VO$_3$. Spectra corresponding to other
compositions are similar to these spectra. Three distinct features
at about 514.5~eV, 516~eV and 517~eV represent the signatures of
V$^{3+}$, V$^{4+}$ and V$^{5+}$ species, respectively.
Accordingly, with a decrease in photon energy, the intensity of
the bulk V$^{4+}$ peak at 516~eV continuously decreases compared
to the intensities of the other two surface related V$^{3+}$ and
V$^{5+}$ features. The ratio of surface and bulk contributions in
2$p$ spectra [${{surface}\over{bulk}} = e^{d/\lambda}-1$, $d$ is
the effective surface depth] provides an estimate of $\lambda$ in
units of $d$. Thus estimated $\lambda/d$ values (see Fig.~1(c))
exhibit linear dependence with $\sqrt{KE}$ for KE~$\geq$~200 eV as
expected from the universal curve \cite{univ-curve}. $\lambda/d$
is found to decrease continuously with KE down to $\sim$16~eV,
presumably due to the presence of various low energy excitations
in these metallic systems. Notably, these experimentally
determined values of $\lambda/d$, are significantly different from
those estimated using Tanuma, Powel and Penn (TPP2M)
formula\cite{tpp2m}.

\begin{figure}
\vspace{-6ex}
 \centerline{\epsfysize=5.0in \epsffile{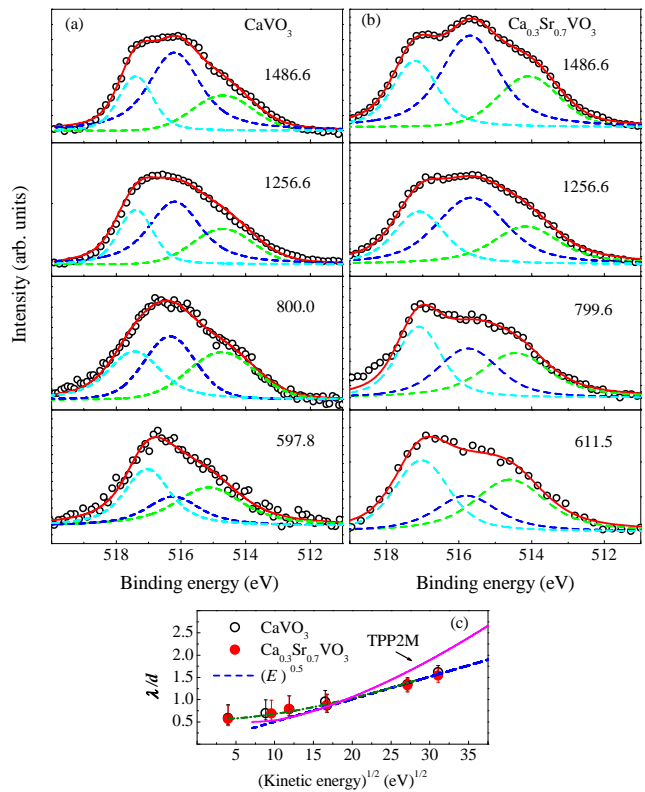}}
 \vspace{-8ex}
 \caption{ V 2$p_{3/2}$ spectra at different photon energies in (a) CaVO$_3$ and (b)
Ca$_{0.3}$Sr$_{0.7}$VO$_3$. Solid lines show the fit using 3 Voigt
functions (dashed lines) representing the signals from V$^{3+}$,
V$^{4+}$ and V$^{5+}$ entities. (c) Estimated $\lambda/d$ vs.
$\sqrt{KE}$. $\lambda/d$ at KE~=~16~eV is obtained from the
valence band analysis \cite{lcvprb2,csv}. Dashed line shows
$\sqrt{KE}$-dependence at higher energies. The solid line
represents the calculated escape depth using Tanuma, Powel and
Penn relations \cite{tpp2m}.}
 \vspace{-4ex}
\end{figure}

The valence band spectra exhibit a series of intriguing results,
not realized so far. For example, the normal emission spectra of
CaVO$_3$ (Fig.~2(a)) with $h\nu$~=~40.8~eV photons from linearly
polarized synchrotron and unpolarized laboratory (He~{\scriptsize
II}) sources are drastically different. The relative intensity of
the incoherent feature at $\sim$1.5~eV is substantially less in
the synchrotron spectrum compared to the He~{\scriptsize II}
spectrum. This, without any change in the photon energy and,
consequently in the electron escape depth or in energy resolution,
specifically kept the same, is curious. Secondly, with a photon
energy of 275~eV from the synchrotron source (Fig.~2(b)), the
ratio of the intensities from the coherent and incoherent features
becomes larger than that observed even with Al~$K\alpha$
radiations (1486.6~eV) on the same sample \cite{csv}. Since the
electron escape depth cannot be larger with $h\nu$~=~275~eV
compared to that at $h\nu$~=~1486.6~eV, it is evident that the
incoherent feature is underestimated in the synchrotron data
compared to the spectra with a laboratory source. The surface
electronic structure contribute primarily to the incoherent
feature. Thus, a suppression of the relative intensity of the
incoherent feature suggests that the photoemission matrix elements
for the surface related states is strongly reduced in the case of
polarized synchrotron radiation compared to unpolarized laboratory
source. While such unexpected discrepancy between the laboratory
source and synchrotron source spectra is already evident in the
results from different groups \cite{morikawa}, this was never
noted in previous studies.

To substantiate these observations, we performed several
experiments on various samples under different experimental
conditions. These experiments reveal a strong dependence of the
coherent-to-incoherent intensity ratio on the angle of incidence
at any fixed photon energy from the synchrotron source as
illustrated by normalizing all the spectra at the incoherent
features of CaVO$_3$ (Fig.~2(b)) and SrVO$_3$ (Fig.~2(c)). This
strong angular dependence of the relative intensities is roughly
independent of the excitation photon energy. For example, a change
in the incidence angle from 45$^o$ to 25$^o$ at $h\nu$~=~275~eV
(Fig. 2(c)) leads to a reduction in the relative intensity of the
coherent feature of SrVO$_3$ by about 14.6\%. This is remarkably
similar to the reduction of $\sim$13.8\% at $h\nu$~=~800~eV
observed for a similar change in angle (Fig. 2(d)). We note that
such an angular dependence of spectral intensities is not observed
with any unpolarized laboratory sources.

Since the angle between the detector and the incident beam in the
experimental set up is fixed at 45$^o$, a change in the incidence
angle also changes the angle of electron emission, thereby being
capable of changing the surface sensitivity. This may provide an
alternate explanation for the change in relative spectral
features. Hence, we investigate the valence band spectra at the
same surface sensitivity, but with different incidence angles by
making the emission angle to be +$\theta$ and -$\theta$ with
respect to the surface normal. A representative case is the
spectra at 55$^o$ and 35$^o$ incidence angles in Fig.~2(b).
Despite same surface sensitivity (emission angle $\pm$10$^o$), the
coherent feature is significantly smaller at 35$^o$ than that at
55$^o$.

In order to ensure that the above results are not artifacts
arising from uncertainties in defining precise emission or
incidence angles due to the unavoidable absence of a well-defined
cleavage plane in such cubic systems, we have simultaneously
probed V~2$p_{3/2}$ core level spectra that provide an internal
measure of surface sensitivity based on distinctly different
surface and bulk spectral features. We chose $h\nu$~=~800~eV for
this purpose, since the core photoelectrons then have the same
kinetic energy as those of valence electrons excited with
$h\nu$~=~275~eV. The core level spectra in Fig.~2(e) at 45$^o$ and
25$^o$ (65$^o$ also shows similar behavior) for {\it both}
CaVO$_3$ and SrVO$_3$ do not exhibit any observable change,
establishing similar surface sensitivity over this range of
angles. This is understandable, as the surface contribution at
normal emission is found to be about 66.4\% ($d/\lambda
\sim$1.09). A change in incidence angle by 20$^o$ leads to a
surface contribution of 68.7\%, representing a change of only
about 2\%. These observations, thus, establish that the spectral
changes with angles in Fig. 2(a)-(d) are not due to a change in
the surface sensitivity, but is indeed related to an intrinsic
reduction in the surface contribution. While the exact origin of
this effect is unclear, the existence of this effect is
unambiguously established by the present experimental results. In
the following we briefly discuss the uniqueness of the surface
electronic structure vis-a-vis that of the bulk, which provide
some clue to understand the observed effects, at least
qualitatively.

\begin{figure}
\vspace{-6ex}
 \centerline{\epsfysize=4.5in \epsffile{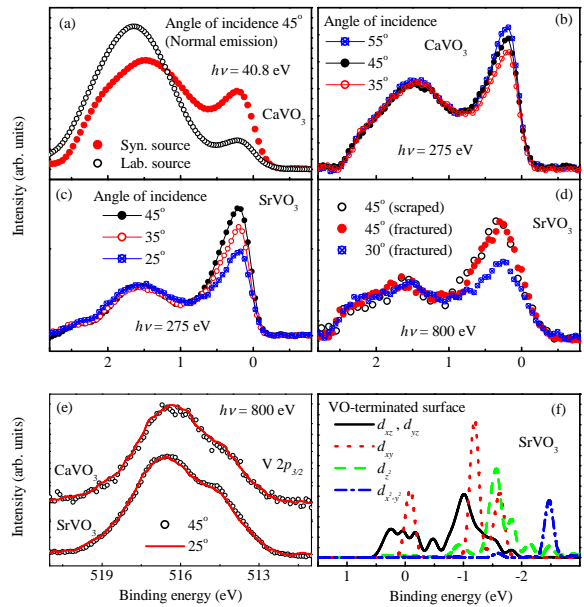}}
\vspace{-18ex}
 \caption{ Valence band spectra of CaVO$_3$ at 40.8~eV
using laboratory and synchrotron sources. Spectra collected at
different incidence angles from (b) CaVO$_3$ and (c) SrVO$_3$
using 275~eV synchrotron radiations, and (d) at 800~eV for
SrVO$_3$. (e) V 2$p_{3/2}$ spectra using 800 eV photons at
different incidence angles. (f) Calculated 3$d$ PDOS of V at
VO-terminated surface. The lineshape of the spectra from scraped
open circles) and fractured (solid circles) shown in (d) is very
similar.}
 \vspace{-4ex}
\end{figure}

The bulk electronic structure of Ca$_{1-x}$Sr$_x$VO$_3$ can be
described essentially in terms of a single $d$ electron
distributed over the triply-degenerate, two-dimensional $t_{2g}$
bands. The crystal symmetry at the surface, however, is expected
to be lowered compared to the octahedral field in the bulk,
leading to a local $D_{4h}$ symmetry. We have confirmed this
expectation by carrying out first principle plane wave
pseudopotential band structure calculation with full geometry
optimization involving a large supercell in a slab configuration.
Resulting partial densities of states (PDOS) at the surface layer
are shown in Fig. 2(f). It is evident in the figure that the
single $d$-electron occupies essentially $d_{zx}$ and $d_{yz}$
bands, while the $d_{xy}$ band is almost empty. Due to the absence
of periodicity along $z$-axis, $d_{zx}$ and $d_{yz}$ bands are
quasi-one dimensional with the $k$-vectors along
(0,0,0)-($\pi$,0,0) and (0,0,0)-(0,$\pi$,0) directions
respectively; interestingly the $d_{xy}$ band, which continues to
be two dimensional at the surface remains unoccupied. The dipole
matrix element, $M_{fi} = \langle\psi_f|A.p|\psi_i\rangle$
($\psi_i$ and $\psi_f$ are the initial and final state
wave-functions) in the expression of photoemission cross section
($I(\epsilon) \propto |M_{fi}|^2f(\epsilon)\delta(\epsilon)$;
$f(\epsilon)$ = Fermi-Dirac distribution function) is a function
of both momentum, $p$ and the vector potential, $A$, and
therefore, the polarization of the incident photons. Thus, the
surface-related band states will have stronger matrix element
effects with polarized synchrotron light compared to the bulk
states due to the lifting of degeneracies at the surface; such
difference will not occur for the unpolarized light source.

\begin{figure}
\vspace{-9ex}
 \centerline{\epsfysize=4.0in \epsffile{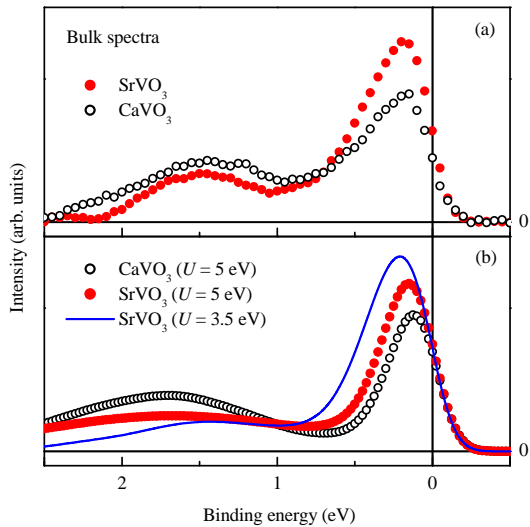}}
\vspace{-12ex}
 \caption{Extracted bulk spectral functions for
SrVO$_3$ (solid circles) and CaVO$_3$ (open circles). (b) DMFT-LDA
results adapted from ref. \cite{andersen}}
 \vspace{-4ex}
\end{figure}

We now extract the bulk contributions, $I^b(\epsilon)$ from the
synchrotron spectra collected at 275~eV and 40.8~eV using the
$d/\lambda$ from Fig.~1(c) in the relation,
$I(\epsilon)=\alpha(1-e^{-d/\lambda})I^s(\epsilon) +
e^{-d/\lambda} I^b(\epsilon)$. $\alpha$, the reduction of the
surface intensity due to polarization, is found to be 0.2$\pm$0.02
by comparing the spectra recorded with $h\nu~=$ 40.8~eV in
Fig.~3(a) \cite{note}. We needed to use a slightly different
$\alpha$ at 275~eV ($\sim$0.1$\pm$0.02) to avoid unphysical
negative intensities. Such a small variation in $\alpha$ is
expected for a fractured surface due to the uncertainties in
defining the surface normal. The extracted $I^b(\epsilon)$ are
shown in Fig.~3(a). Interestingly, $I^b(\epsilon)$ of CaVO$_3$ and
SrVO$_3$ are significantly different exhibiting the relative
intensity of the coherent feature compared to the incoherent one
is distinctly larger for SrVO$_3$; this is indeed what should be
expected from the fact that the bandwidth in SrVO$_3$ is
significantly larger. This expectation appears to be fully
justified by the results of a most sophisticated {\em ab initio}
DMFT~+~LDA calculations reported recently \cite{andersen}, which
is adopted in Fig. 3(b) for U~=~5~eV. These calculated spectra
with a distinctly larger relative intensity of the coherent
feature in SrVO$_3$ compared to CaVO$_3$, exhibit similar
differences in the electronic structures of these two compounds,
leading to an unified understanding of this interesting class of
compounds and removing the latest puzzling aspects of its reported
electronic structure. We also stress the point that the present
results are consistent with the significant enhancement in
$\gamma$ of CaVO$_3$ compared to SrVO$_3$ \cite{csv}; this is
another experimental fact that would be difficult to reconcile
with the earlier reported identical electronic structure for the
two compounds. It is, however, to be noted that though there is a
good qualitative agreement between the present experimental
results and the independent theoretical ones, Fig.~3 also
underlines the same interestingly and possibly significant
quantitative discrepancies between theory and experiment.
Specifically, the incoherent features appear at higher energies in
the calculations compared to experimental results, though the
relative intensities are reasonably well described. When the
energy position of the incoherent feature is brought to better
agreement by a reduction in $U$, the calculated relative intensity
becomes unreasonably low, as illustrated in Fig.~3(b) for SrVO$_3$
with $U$~=~3.5~eV. This is strongly reminiscent of the Ni
satellite problem, where a simultaneous quantitatively accurate
description of both the satellite (incoherent feature) intensity
and the energy position has remained elusive. The present results
suggest that this may be a more general problem related to the
description of strongly correlated transition metal based systems
in terms of the simple Hubbard model and require further
theoretical inputs.

Keeping in mind the above-mentioned caveat, the present results
still clearly establish that the linear polarization of
synchrotron radiation plays a key role in determining the spectral
lineshape in these systems due to strong matrix element effects.
The experimentally-determined bulk spectra provide an
understanding of the electronic structure in
Ca$_{1-x}$Sr$_x$VO$_3$, consistent with experimental $\gamma$
values and the geometrical/structural trends across the series,
thereby resolving the puzzle concerning the structure-property
relationship in this interesting class of compounds.

The authors acknowledge financial support from ICTP-Elettra, Italy
and DST, Govt. of India. U.M. acknowledges the financial support
from CSIR, Govt. of India.


\begin{thebibliography}{99}
\bibitem{note} $I^s(\epsilon)$ in CaVO$_3$ essentially contributes in the
incoherent feature intensity \cite{lcvprb2,csv}. Thus, the
difference in incoherent feature intensity relative to coherent
feature provides an estimation of $\alpha$.
\end{thebibliography}
\end{document}